\documentclass[letter]{aa} 
\usepackage{natbib,twoopt}
\usepackage[breaklinks=true]{hyperref} 
\usepackage{comment}
\usepackage{parskip}
\usepackage{gensymb}
\usepackage{subcaption} 
\usepackage{graphicx}
\usepackage{txfonts}

\bibpunct{(}{)}{;}{a}{}{,}             
\newcommand{\DEL}[1]{}

\begin{document} 
\title{
General Relativistic effects and the NIR variability of Sgr A* II}
\subtitle{A systematic approach to temporal asymmetry}

\author{
Sebastiano D. von Fellenberg\inst{1}
      \and Gunther Witzel\inst{1}
      \and Michi Bauboeck \inst{2}
      \and Hui-Hsuan Chung\inst{1}
      \and Nicola Marchili\inst{3}
      \and Greg Martinez \inst{5}
      \and Matteo Sadun-Bordoni \inst{4}
      \and Guillaume Bourdarot \inst{4}
      \and Tuan Do\inst{5}
      \and Antonia Drescher \inst{4}
      \and Giovanni Fazio \inst{6}
      \and Frank Eisenhauer \inst{4}
      \and Reinhard Genzel \inst{4}
      \and Stefan Gillessen \inst{4}
      \and Joseph L. Hora \inst{6}
      \and Felix Mang \inst{4,7}
      \and Thomas Ott \inst{4}
      \and Howard A. Smith \inst{6}
      \and Eduardo Ros \inst{1}
      \and Diogo C. Ribeiro \inst{4}
      \and Felix Widmann \inst{4}
      \and S. P. Willner \inst{6}
      \and J. Anton Zensus \inst{1}
      }

\institute{
        Max-Planck-Institute für Radioastronomie, 
        Auf dem H{\"u}gel 69, D-53121 Bonn, Germany 
        \and University of Illinois, Urbana-Champagne, USA
        \and Italian ALMA Regional Centre, INAF-Istituto di Radioastronomia, Via P. Gobetti 101, I-40129 Bologna, Italy 
        \and Max Planck Institut for Extraterrestrial Physics, D-86 Garching bei Muenchen, Germany
        \and Physics and Astronomy Department, University of California, Los Angeles, CA 90095-1547, USA; tdo@astro.ucla.edu
        \and Center for Astrophysics $|$ Harvard \& Smithsonian, 60 Garden Street, Cambridge, MA 02138, USA
        \and Technical University of Munich, 85747 Garching, Germany
        }

\date{Created \today ; accepted }

\abstract{A systematic study, based on the third-moment structure function, of Sgr~A*'s variability finds an exponential rise  time  $\tau_{1,\rm{obs}}=14.8^{+0.4}_{-1.5}~\mathrm{minutes}$ and decay time  $\tau_{2,\rm{obs}}=13.1^{+1.3}_{-1.4}~\mathrm{minutes}$. This symmetry of the flux-density variability is consistent with earlier work, and we interpret it as caused by the dominance of Doppler boosting, as opposed to gravitational lensing, in Sgr~A*'s light curve. A relativistic, semi-physical model of Sgr~A* confirms an inclination angle $i \leq 45 \degree$. The model also shows that the emission of the intrinsic radiative process can have some  asymmetry even though the observed emission does not. The third-moment structure function, which is a measure of the skewness of the light-curve increments, may be a useful summary statistic in other contexts of astronomy because it senses only temporal asymmetry, i.e., it averages to zero for any temporally symmetric signal.}
 
   

   \maketitle
%
\section{Introduction}
\label{sec: intro}
Ever since the detection of the near-infrared (NIR) and X-ray counterpart \citep{Baganoff2001_flare,Genzel2003} of the radio source Sgr~A*, the source's variability at all wavelengths has been the subject of intense study. 
Sgr~A*'s NIR light curve shows occasional bright outbursts, typically phenomenologically referred to as flares. A NIR flare is always present when an X-ray outburst occurs \citep{Eckart2008}, but the converse is not true: only a fraction of bright NIR flares are  accompanied by an X-ray flare \citep[e.g.,][]{Dodds-Eden2009}. Even when flares occur in both bands, the respective flux levels are not highly correlated \citep[e.g.,][]{GravityCollaboration2021_xrayflare}.
The spectral shape of NIR-to X-ray flares has now been  established to be rising in the NIR as $\nu L_\nu \propto\nu^{ +0.5}$ and falling in the X-ray as $\nu L_\nu \propto \nu^{ -0.5}$ \citep{Hornstein2007, Yusef-Zadeh2009, Ponti2017, GravityCollaboration2021_xrayflare}. 

Modern  interferometric  observatories  can  detect  Sgr~A*  at  all times \citep{GravityCollaboration2020flux}, which reveals the Sgr~A* is a  constantly  variable  source even  in  the  absence  of higher  flux  peaks  (flares).  This  NIR variability  has  been  studied  in detail: the source shows stochastic, red-noise-like behavior  akin  to  many  other  compact  objects \citep[e.g.,][]{Do2009}.  The power spectrum shows no features and breaks into uncorrelated  white  noise  on  time scales of ${\approx} 150 ~\mathrm{minutes}$ \citep[e.g.,][]{Witzel2012, Witzel2018}. The RMS--flux relation is linear with no  significant difference in variability for bright and faint parts of the light curve, again very similar to other compact objects \citep{Witzel2012, GravityCollaboration2020flux, Weldon2023}. On the other hand, the NIR flux distribution is log-right-skewed, i.e., it shows a tail with respect to a log-normal distribution \citep{Dodds-Eden2009, Witzel2012, GravityCollaboration2020flux}. This tail is typically used to motivate the concept of flares as events in the accretion flow, contrasting with the quiescent condition.

If flares are modeled as distinct events, two types of one-zone models can describe the data. One type peaks at submillimeter (submm) wavelengths and produces NIR and X-ray flux through synchrotron-self-Compton (SSC) emission \citep[e.g.,][]{Eckart2008, Yusef-Zadeh2009, Witzel2021}. In the other type, both the NIR and X-ray emission are caused entirely by synchrotron emission, and the resulting but unobserved SSC emission peaks at gamma-ray energies. Models of the first type require enormous electron densities to obtain sufficient SSC flux to reproduce the X-ray spectral index and therefore are often disfavored \citep{Dodds-Eden2011, Ponti2017, GravityCollaboration2021_xrayflare}.

An additional observational input to models is that the astrometry of NIR flares can be tracked. Observed flares show close-to-circular proper motions in the sky, the so-called orbital motions \citep{GravityCollaboration2018_orbital, GravityCollaboration2020_orbital}. In addition, NIR flares show characteristic loop-like swings in linear polarization \citep{GravityCollaboration2020_polariflares, GRAVITY_collab2023_polariflares}. These NIR observations were  corroborated by EHT--ALMA observations \citep{Wielgus2022,eht_sgra_I}, which showed that a similar polarization loop was detected in the  $230~\mathrm{GHz}$ radio light curve, incident right after a bright X-ray flare. The astrometric behavior  supports the notion of an event-like character of flares, where flares are interpreted as orbiting ``hot spots'' in the accretion flow.

If the flare motion is interpreted as an accelerating point-particle orbiting Sgr~A*, orbital parameters such as the radius and the viewing angle can be constrained. Both the NIR data and the submm polarization loop  \citep{Wielgus2022} are consistent with a face-on inclination. The Sgr~A* EHT image also indicates that the accretion flow is viewed face-on \citep{eht_sgra_I, eht_sgra_II, eht_paper_III, eht_paper_IV, eht_paper_V, eht_paper_VI}. Thus there are four independent indications that Sgr A*'s accretion flow is oriented face-on: NIR astrometry of flares, NIR polarization of flares, radio polarization of the light curve after an NIR flare, and the EHT image of Sgr~A*. 

Paper~I \citep{vonFellenberg2023_sgra} analyzed light curves obtained in the NIR ($4.5~\mathrm{micron}$, Spitzer) and in the X-ray (2--8{keV}, Chandra, XMM-Newton). On average, Sgr~A* flares are well approximated by a double-sided, symmetric, exponential profile with a rise and decay time $\tau\approx 15~\mathrm{min}$.
For a fully relativistic and ray-traced orbiting hot-spot model \citep{GravityCollaboration2020_orbital}, such a profile can only be explained if gravitational lensing in the light curve is not important, i.e., if the orbits are viewed face-on.  In this picture, the symmetry of the average flare shape is explained by dominant Doppler-boosting as the hot spot orbits. When averaged over several flares, the different initial orbital positions lead to symmetric Doppler amplification of the underlying emission. 
The symmetric flare shape is a fifth, independent constraint on the viewing angle of Sgr~A*. This constraint uses only the variability and the assumption of moving hot spots close to Sgr~A*. This constraint therefore applies to all hot-spot models, as the variability induced by such a moving hot spot is valid for any radiative model of the underlying emission mechanism, as any close-to-circular motion close to Sgr~A* will be lensed and boosted. 

A limitation in paper I was the arbitrary selection of flares, their flux normalization, and the shifting and adding applied.
This paper generalizes the analysis by first establishing summary statistics to measure asymmetry based on flux differences. Section~\ref{sec:sf} introduces a statistical model to measure asymmetry in a stochastic process, and Section~\ref{sec:lc} applies the method to fit the observed Sgr~A* light curves (Section~\ref{sec: data}).  Section~\ref{sec:gr} introduces a semi-physical general-relativistic model and uses it to derive intrinsic Sgr~A* properties consistent with the observed light curves, and Section~\ref{sec: discussion} puts the results in context. All fits to data make use of the approximate Bayesian Computation (ABC) method as described in \autoref{sec:parameter_estimation}.

\section{Data}
\label{sec: data}


Our data consist of eight NIR observations by \textit{Spitzer}/IRAC, each $\sim$24 continuous hours, observed from 2013 to 2017. The timing resolution is 8.4 seconds, which we re-binned to a cadence of 1 minute.  Throughout this paper we will use data that are not corrected for extinction, i.e., we will use observed flux densities. This choice makes no difference to our results except for overall normalization of light curves. The data are public \citep{Witzel2018, Witzel2021}, and the calibration approach was explained by \citet{Hora2014}.  As those authors explain, the flux density zeropoint is unknown, but that makes no difference for this paper.

\section{Light curves and structure functions}
\label{sec:sf}
\DEL{
In the following (\autoref{sec:differential_fluxes}), we will derive the framework used to constrain the asymmetry of Sgr A*. In particular, we will make use of summary statistics, which, when suitably chosen, can represent the data well.

The summary statistics will be used to determine the statistical properties of Sgr A*'s NIR light curve. We will construct different summary statistics that will us allow to independently constrain Sgr A*'s flux distribution, its auto-correlation, and the asymmetry.

Lastly, we introduce a generic model to generate random realizations of Sgr A*'s light curves. We will use the moving average (MA) model (\autoref{sec: ma_model}, \cite{Priestley1988,Scargle1981,Scargle2020}). The moving average model is chosen over other, similarly generic models such as autoregressive (AR) models or Gaussian processes \citep{some}, as it allows for an intuitive definition of asymmetry, namely the asymmetry of the system response function. 

We will optimize the model parameters, by \textit{fitting} the model structure functions to the structure functions of the observed \textit{Spitzer} data. The fits will be derived using the well-established method of \textit{Bayesian Approximate Computation} (BAC) \citep{some} implemented in the code \textsc{pyabc} \citep{klinger2018pyabc, schaelte2022pyabc}, which is well established in the statistical literature and has successfully been applied to Sgr A* light curves \citep[e.g.,][]{Witzel2018}. 
}

\begin{figure*}
    \centering
    \includegraphics[width=0.985\textwidth]{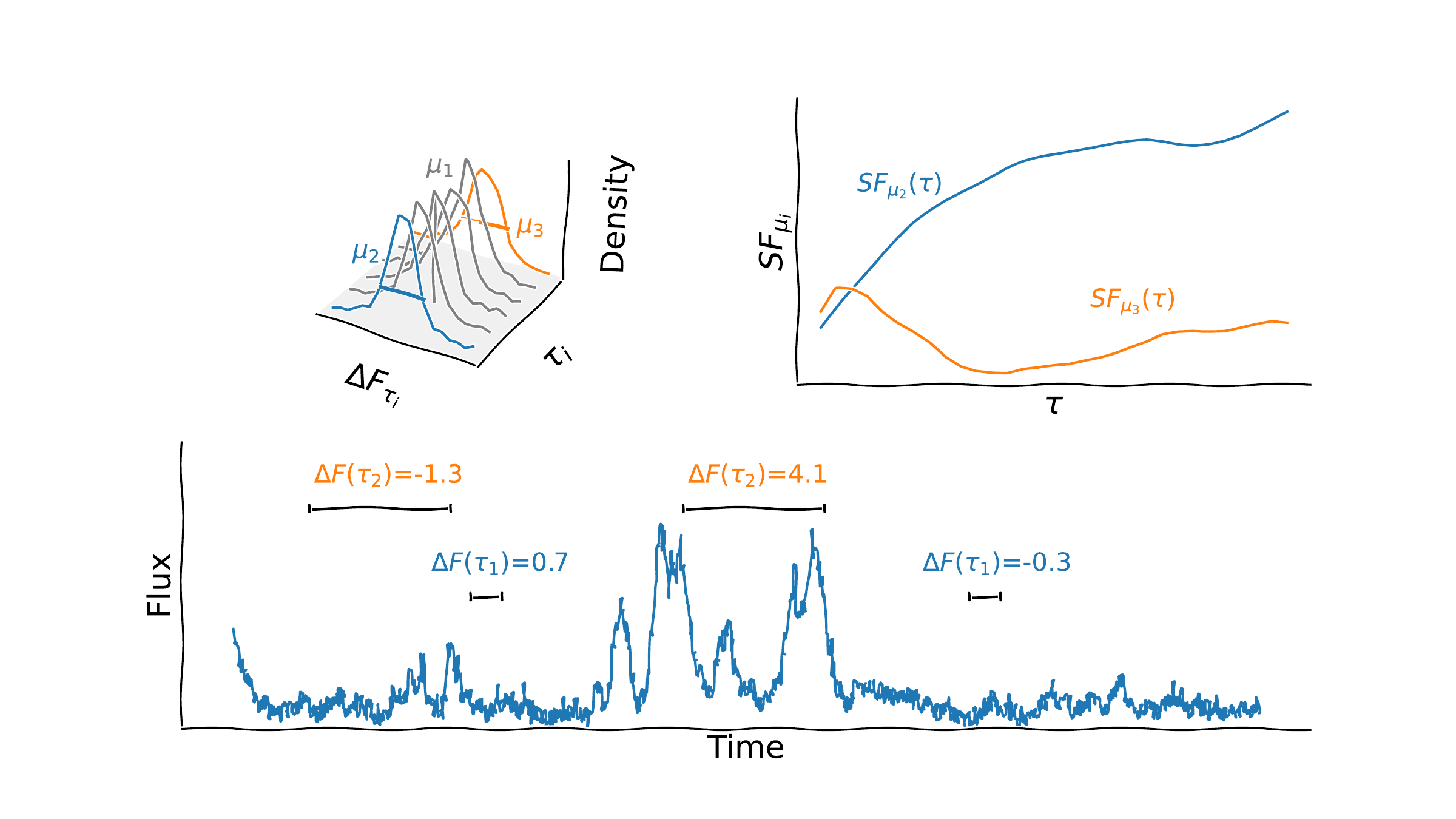}
    \caption{Cartoon showing the concept of moment-based summary statistics for light-curve analysis. The bottom panel shows an example light curve, the orange and blue annotation indicate the flux differences for two given time lags $\tau_1$ and $\tau_2$. The top panel shows the histogram of flux differences for every time lag $\tau_i$. Each histogram has its respective statistical moments:  mean ($\mu_1$),  variance ($\mu_2$), and  skewness ($\mu_3$). At upper right, variance and skewness are plotted as functions of time lag.}
    \label{fig:illu_flux_differences}
\end{figure*}

\label{sec:differential_fluxes}
To characterize the statistical properties of light curves, this paper uses flux-density differences of  pairs  of points separated by a given time lag $\tau$. This is a common approach in  astronomy  \citep[e.g.,][]{Simonetti_1985}. 
\autoref{fig:illu_flux_differences} illustrates the underlying concept: for each time-lag $\tau$, all available flux pairs in a light curve are measured and histogrammed. Then the statistical moments of each histogram, here the mean ($\mu_1$), the variance ($\mu_2$), and the (Fischer) skewness ($\mu_3$), are calculated. The value of any of these moments as a  function of the time lag $\tau_1$ is the structure function:
\begin{equation}
    \mathrm{SF}(\tau, \mu_i) = c \times \mu_i(\tau)\quad,
\end{equation}
where $c$ is a normalization constant. 
For practical reasons, and to increase the fidelity of the estimate, time lags are grouped in bins $\tau_i$ of suitable width in $\tau$. 

\subsection{The second-moment structure function}
The vast majority of astronomical works  use the second-moment (variance) structure function $\mu_2 = \sigma^2$. Different normalizations have been used in the literature,  but a standard formula for the unnormalized SF \citep[$\rm{SF_{\mu_2}}$, e.g., ][]{Simonetti_1985} is
\begin{equation}
\label{eq: SF}
\mathrm{SF'}_{\mu_2}(\tau_{i}) = \frac{1}{N_{i}}\sum_{t_{j}, t_{k}}[F(t_{j}) - F(t_{k})]^{2}~ ,~ \tau_{i} < (t_{j} - t_{k}) \leq \tau_{i+1}~.
\end{equation}
Here, $N$ is the number of data pairs in each time-lag bin $i$, and $F(t_j) - F(t_k)$ is the difference in flux density between times $t_j$ and $t_k$. 

For this work, we will normalize the structure-function to the interval $\mathrm{SF}(\tau) \in [0, 1]$:
\begin{equation}
    \mathrm{SF}_{\mu_2}(\tau_{i}) = \frac{1}{4 N_{i} \sigma^2(F)} \mathrm{SF'}_{\mu_2}(\tau_i)\quad,
\end{equation}
which is identical to definition in \autoref{eq: SF} up to the factor $1/4\sigma^2(F)$, where $\sigma^2(F)$ is the variance of the light curve.
This choice of normalization {renders the $\mathrm{SF_{\mu_2}}$} dimensionless, does not depend on the absolute flux levels, and measures only the correlation in the data. This can be understood when a mathematical property of the second moment structure function is taken into account. The structure function forms a Fourier pair of the power spectrum of a light curve (a consequence of the Parceval's theorem \citep{dechenes_1799}), and this normalization removes the absolute value of the power spectrum. Therefore $\rm{SF_{\mu_2}}$ is sensitive only to the slope of the power spectrum.

\subsection{The third-moment structure function}
\label{sec: Skew}
Following the logic of $\rm{SF_{\mu_2}}$, the third moment of the flux density differences, the skewness $\mu_3$ defines
the unnormalized third-moment structure function
\begin{equation}
    \label{eq: skew}
    \mathrm{SF'}_{\mu_3}(\tau_i) = \frac{1}{N_{i}} \sum_{t_{j}, t_{k}}[F(t_{j}) - F(t_{k})]^{3}
\end{equation}
with the same symbols as in \autoref{eq: SF}. To obtain a unit-free normalization in the interval $\mathrm{SF}_{\mu_3}(\tau) \in [0,1]$ we apply
\begin{equation}
    \label{eq:sf3_norm}
    \mathrm{SF}_{\mu_3}(\tau_i) = \dfrac{\sqrt{N_{\rm{tot}}}}{\mu_3(F)} \times \dfrac{\rm{SF'_{\mu_3}(\tau_i)}}{ \mathrm{(SF'}_{\mu_2}(\tau_i))^{3/2}}~,
\end{equation}
where  $\sigma^2(F)$ denotes the variance of the light curve, and $\mu_3(F)$ denotes its  Fischer skewness \citep[e.g.,][]{Joanes1998} defined as
\begin{equation}
    \label{eqn:fischer_skewness}
    \mu_3 = \dfrac{E[(X-\mu)^3]}{E[(X-\mu)^2]^{3/2}}\quad,
\end{equation}
where $E$ stands for the expectation value.
As for  $\rm{SF_{\mu_2}(\tau)}$, this dimensionless normalization ensures that the third moment structure function is sensitive only to the asymmetry in the flux differences and is not dependent on the slope of the power spectrum. 

\begin{figure*}
    \centering
    \includegraphics[width=0.99\textwidth, trim=0 0 12.5cm 0, clip]{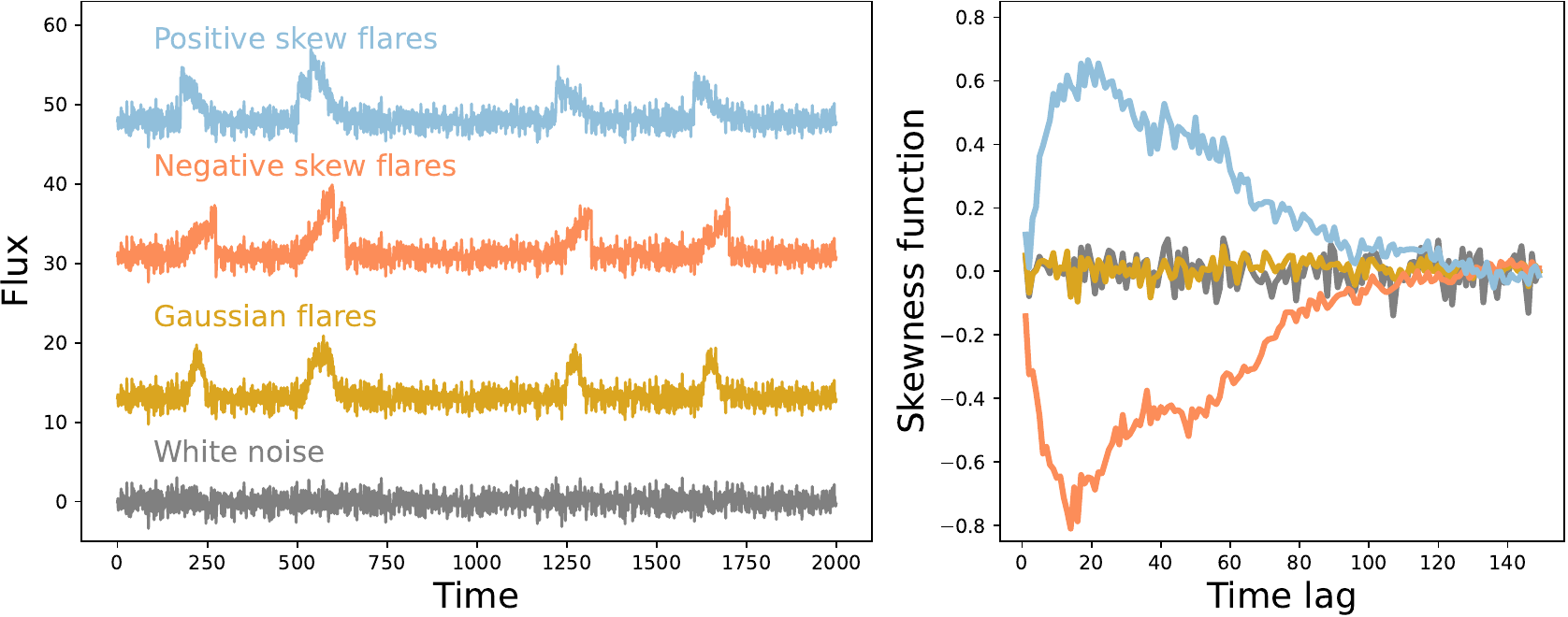}
    \caption{Illustration of the properties of $\rm{SF_{\mu_3}(\tau)}$.  The left panel shows four artificial light curves of different types as labeled, and the right panel shows their skewness functions defined by \autoref{eq:sf3_norm}.}
    \label{fig:SF3_example}
\end{figure*}

The third-moment structure functions possess several useful properties illustrated in \autoref{fig:SF3_example}.\footnote{There have been attempts to use modifications of the structure-function to measure asymmetry in light curves \citep[e.g.,][]{Kawaguchi_1998, Hawkins_2002, Chen_2015}, non of which probe temporal asymmetry directly. \cite{Shishov2005} used an asymmetry function to probe the flux density distributions of pulsars for interstellar scintillation, which, however, does not probe temporal asymmetry.}
The third moment structure function shows a negative sign for positively skewed flares  and a negative sign for negatively skewed flares. For symmetrical flares, the skewness function is approximately zero for all time lags. Similarly, white noise, which is added to all curves, is approximately zero{, on average, in any one realization}. 

Thus the $\rm{SF_{\mu_3}(\tau)}$ is sensitive \textit{only} to asymmetric signals, and white noise averages to zero when there are enough data points. 
These properties make the third-moment structure function a versatile metric for astronomical signals that are \textit{intrinsically asymmetric} (e.g. QPEs \cite{Arcodia2021}, eclipsing binaries \cite{Gautam2024}, ...). 

\section{Quantifying asymmetry in light curves}
\label{sec:lc}
\subsection{An intuitive overview of the method}
While $\mu_3$ (\autoref{eqn:fischer_skewness}) easily quantifies asymmetry in a distribution of measurements (or random numbers), it is not immediately apparent how asymmetry can be understood for a time series.
For a singular \textit{event}, asymmetry can easily be defined by whether the left and right flanks around the event's peak are identical (symmetric) or not (asymmetric). While an event is not a probability distribution, the mathematical formalism to calculate the third moment can be applied, and positive or negative skew will translate to positive or negative $\rm{SF_{\mu_3}(\tau)}$ (\autoref{eq:sf3_norm}) values. 
What works for a single peak must also work for repeating, isolated events, and asymmetry is easily probed in that case by the sign of $\rm{SF_{\mu_3}(\tau)}$, as illustrated in \autoref{fig:SF3_example}. {We discuss a proper definition of symmetry under time reversal in \autoref{ap:proof} and proof that in the case of symmetry the third moment structure function is zero}.
The sign of the $\rm{SF_{\mu_3}(\tau)}$ also probes asymmetry for different light curve types, including when individual flares are less isolated and overlap or when no easy definition of an individual event is apparent. 

The difficult part of using $\rm{SF_{\mu_3}(\tau)}$ to measure asymmetry in a light curve is to quantify the asymmetry and its statistical significance. Any light curve can be split up into segments, some of which will, by chance, show positive or negative values of $\rm{SF_{\mu_3}(\tau)}$. As for any statistic, with fewer data pairs, the moment $\mu_3$ of the flux difference distribution becomes less well determined. 
The following sections establish a method to quantify the significance of asymmetry in the light curve based on the moving-average (MA) model for time series. This model allows generic symmetric and asymmetric light-curve generation as well as an easy definition of the magnitude of asymmetry and its significance. This approach has largely been inspired by the discussion of \cite{Scargle2020}, who discussed the utility of MA models for astronomy.

\subsection{Moving average model}
\label{sec: ma_model}

The MA model is a generic way of generating a stationary light curve \citep[e.g.,][]{Priestley1988,Scargle1981,Scargle2020}. Following the notation of \cite{Scargle2020}, a light-curve point $X(n)$ at time step $n$ is given by
\begin{equation}
\label{eq: ma_process}
    X(n) = \sum_k c_k R(n-k) + D(n)\quad,
\end{equation}
where $c_k$ are the MA coefficients called the \textit{impulse response} of the process. $R(n)$ are uncorrelated random numbers at each time step $k$, typically called the \textit{innovation}, and $D(n)$ is a deterministic function. Effectively, this can be expressed as the convolution of the random process $R(n)$ with the impulse response $C$:
\begin{equation}
    X(n) = R(n) * C + D(n)\quad.
\end{equation}
It can be shown that, for a suitable choice of $c_k$ and $R(n)$, the MA model is capable of representing \textit{any} stationary light curve \citep[{Wold} theorem:][]{Wold1964, Scargle2020}. MA is therefore suitable to generate mock light curves depending only on the choice of impulse response, the chosen distribution of random numbers, and the luck of the actual random numbers generated. 

In generating mock light curves, the random numbers $R(n)$ need not be drawn from a uniform distribution. Again following \cite{Scargle2020}, we chose a $\beta$-distribution, i.e., a random variable $x$, drawn from uniformly distributed random numbers $\mathcal{U} \in [0,1]$, generates 
\begin{equation}
\label{eq: innov}
R(x)=x^{\alpha}~,\quad \alpha > 0\quad.
\end{equation}
Choosing $\alpha=0$ gives white noise.
\autoref{fig: ma_lc_generation} demonstrates the effect of larger values of $\alpha$:  $\alpha=10$ leads to a non-uniform, but not highly skewed, distribution of random values $R$. This, when convolved with the kernel $C$, leads to a light curve where individual events are hardly stand out. 
Higher values of $\alpha$ lead to more skewed innovations and more distinct individual events in the light curve. 

The choice of kernel $C$ is arbitrary, and while it can be measured from the data, it is difficult to find a \textit{good} kernel a~priori. Paper~I showed that flares of Sgr~A* are well represented, at least on average, by a double exponential function. In particular, Paper~I used a PCA decomposition of stacked flares to obtain a de-noised kernel, derive a $\tau_{+/-}\sim 15 ~\mathrm{minutes}$, and show that PCA can differentiate different kernel shapes for a broad range of MA model parameters. This motivates our choice of a doubled-sided exponential parameterized by $\tau_1$ and $\tau_2$ for the impulse response:
\begin{equation}
  C(t, \tau_1, \tau_2) =    \begin{cases}
      e^{|t|/\tau_{1}} & t\leq 0\\
      e^{-|t|/\tau_{2}} & t>0~. \\
    \end{cases}      
    \label{eq: eq_kernel}
\end{equation}
This choice has the advantage that 1) the MA model can be fit to the light-curve data using the structure functions discussed in \autoref{sec:differential_fluxes}, and 2) the asymmetry in the light curve can be expressed as $\Delta \tau = \tau_1 - \tau_2$. 
The uncertainties of the values $\tau_1$, $\tau_2$, and $\Delta \tau$ can be derived from the posteriors of the model fit.

\begin{figure}
    \centering
    \begin{subfigure}[b]{0.485\textwidth}
        \centering
        \includegraphics[width=\textwidth]{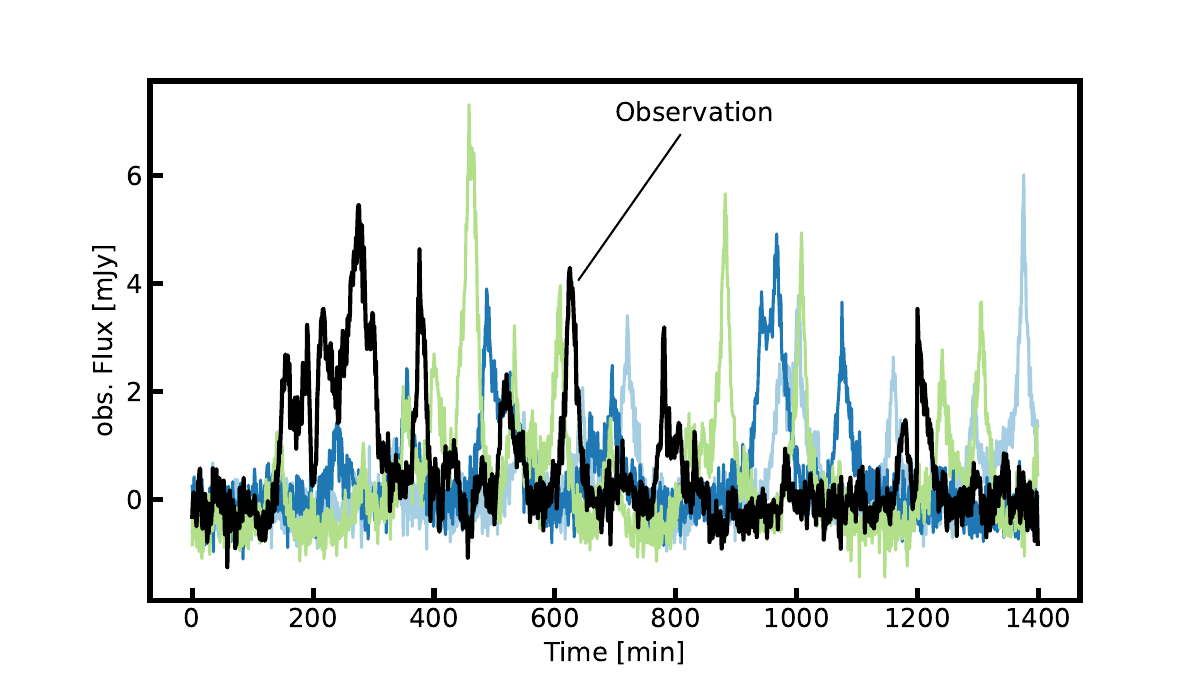}
    \end{subfigure}
    \\
    \begin{subfigure}[b]{0.485\textwidth}
        \centering
        \includegraphics[width=\textwidth,clip=true,trim=0 0 0 30]{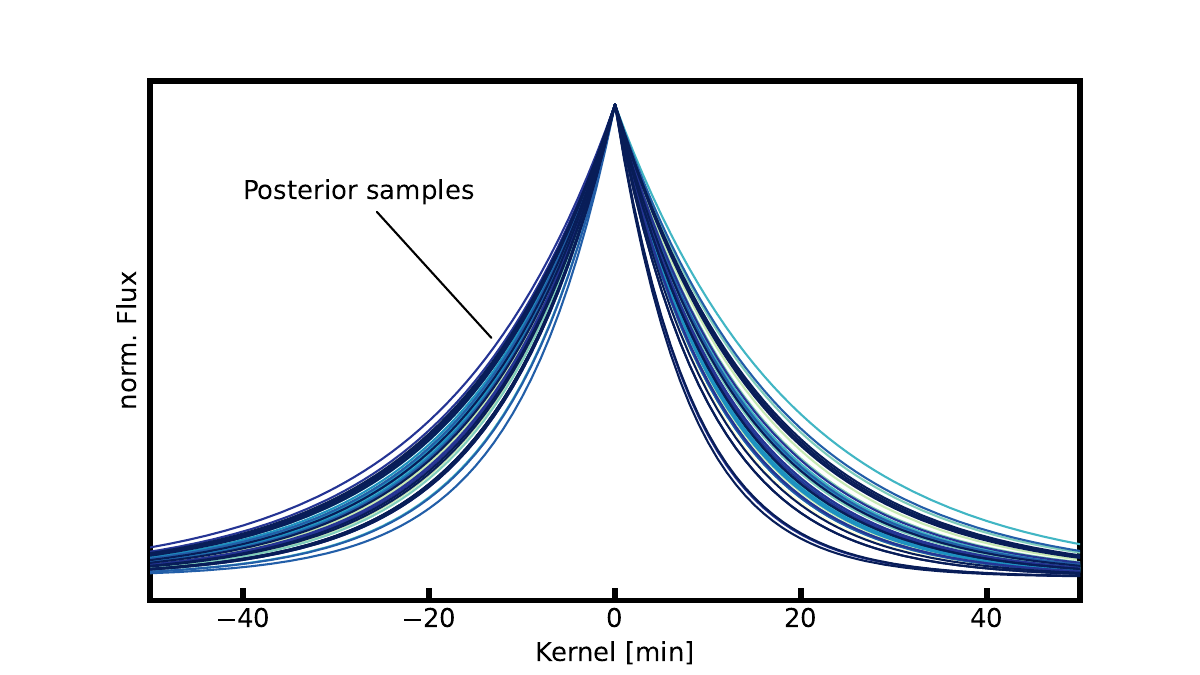}
    \end{subfigure}
    \caption{\textit{Top:} one real and three mock light curves. Colored lines show mock light curves generated with the MA model and best-fit posteriors. The black line shows a segment of an observed Sgr~A* light curve. \textit{Bottom:} kernel functions constructed using $\tau_{1,2}$ drawn from the MA-model posterior.}
    \label{fig:ma_posterior}
\end{figure}

\subsection{Parameter estimation by ABC\label{sec:parameter_estimation}}

To find what values of $\tau_1$ and $\tau_2$ best fit the observed light curves, we used the approximate Bayesian computation (ABC) approach. The model requires two additional parameters, $\alpha$ and an overall flux-density normalization $f_{\rm scale}$. Specifically,
\begin{equation}
\begin{aligned}
    F&_{\rm{model}}(t; \tau_{1,\rm{obs}}, \tau_{2,\rm{obs}}, f_{\rm{scale}}, \alpha) =  \\ 
    &f_{\rm{scale}} \cdot C_{\rm{obs}}(\tau_{1,\rm{obs}}, \tau_{2,\rm{obs}}) * \mathcal{U}(t)^{\alpha} + \mathcal{N}(\sigma_{\rm{obs.}}) - {M}(F)~,
    \label{eq:model}
\end{aligned}
\end{equation}

where $\mathcal{U}^\alpha$ is the innovation {shown here} {convolved with} $C_{\rm{obs}}(\tau_{1,\rm{obs}}, \tau_{1,\rm{obs}})$ the observed impulse response as defined above, $f_{\rm{scale}}$ is the normalization constant. The {amplitude} {of the} measured noise in the rebinned light curves $\sigma_{\rm{obs}}$ and process median ${M}(F)$ are known in advance and need not be solved for, but each model light curve requires a random realization of the measured noise $\mathcal{N}(\sigma_{\rm{obs.}})$. 

The model parameters were constrained using the \textsc{pyabc} code \citep{klinger2018pyabc, schaelte2022pyabc}, which minimizes the distance between the summary statistics of the data and of the model. Our choice of unit-free normalization for both structure functions causes a selective parameter sensitivity {in the numerical problem}
\begin{itemize}
    \item $\rm{SF_{\mu_2}(\tau)}$ mostly constrains the average width of the kernel $\langle\tau\rangle\equiv (\tau_1+ \tau_2$)/2;
    \item $\rm{SF_{\mu_3}(\tau)}$ mostly constrains the impulse response difference $\Delta \tau \equiv  (\tau_2 - \tau_1)$.
\end{itemize}
To constrain $\alpha$ and $f_{\rm{scale}}$, we included a third summary statistic, namely the KS-test, to estimate the difference between model and data flux distributions. \autoref{tab: model_parameter} gives an overview of model parameters and the chosen priors. 

\subsection{Light-curve asymmetry results}
Sgr A*'s \textit{observed} light curve in the NIR can be successfully modeled with an almost symmetric kernel, with an average rise and decay time $\tau\approx 15~\mathrm{minutes}$. {\autoref{tab: model_parameter} lists the model parameters, their uninformative priors, and the resulting posteriors.} \autoref{fig:ma_posterior} illustrates example light curves and   impulse responses, which are all nearly symmetric. 
The result is consistent with that of paper~I, where only the shape of the average flares in the light curve was modeled. 
In particular, the constraints on the rise and fall time are
\begin{equation}
\centering
\begin{aligned}
    \tau_{1,\rm{obs}}=14.8^{+0.4}_{-1.5}~,~
    \tau_{2,\rm{obs}}=13.1^{+1.3}_{-1.4}.
\end{aligned}
\end{equation}
and there is no significant difference between the rise and fall times of Sgr~A*'s observed impulse response $C_{\rm{obs}}$.

\begin{figure}
    \centering
    \includegraphics[width=0.485\textwidth]{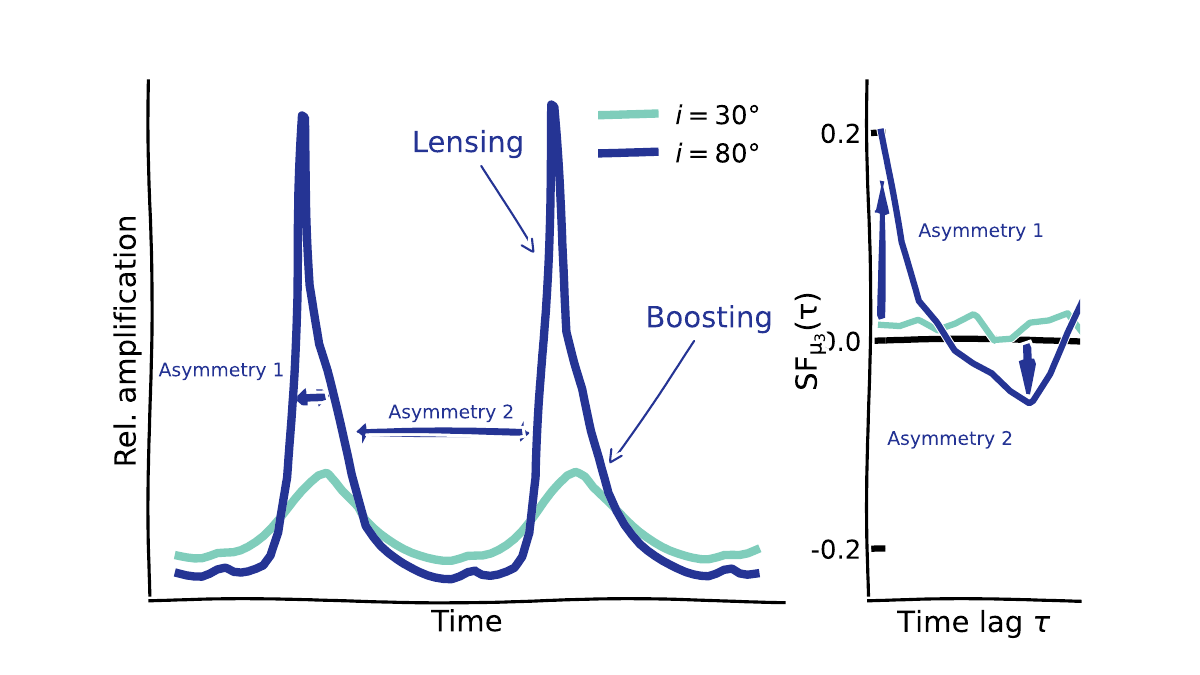}
    \caption{relativistic magnification in Sgr~A*'s light curve. The left graph shows the magnification kernel computed from a library of ray-traced images \cite{GravityCollaboration2020_orbital} for two different inclination angles as indicated by the line colors. The time interval plotted is two orbits.  Both lensing and boosting peak at orbit phase near $1.5\pi$ \citep{Hamaus2009}. The right graph shows 
    $\rm{SF_{\mu_3}(\tau)}$ computed for each magnification curve using \autoref{eq:gr_model}.}
    \label{fig:example_gr}
\end{figure}

\section{General relativistic effects}
\label{sec:gr}
\subsection{A semi-physical relativistic Sgr~A* model}
\DEL{
In this section, expand the generic time series model from before to a physical model of Sgr A*. The model is kept as simple as possible. Our specific aim is to determine the impact of relativistic effects on the variability of Sgr A*. 
}

The \textit{intrinsic} variability of Sgr~A* is modified by relativistic effects{, such as Doppler boosting and relativistic lensing}. To explore these, we use the so-called ``orbiting hot-spot model'' in which bright flares are caused by a localized zone moving in the accretion flow. The flares are assumed to move on bound orbits around the black hole and will be boosted and lensed depending on the observer viewing angle and the position in the orbit. 
Magnifying effects comprise both Doppler boosting and lensing.
Both  are inclination-angle dependent and are minimum for face-on orbits. Increasing the inclination angle 
increases both effects but by differing amounts. At $i < 30\degree$, lensing is negligible, but for  $i > 50\degree$, lensing dominates. In addition, lensing is ``faster,'' leading to a distinct magnification peak which is broadened by the ``slower'' boosting contribution. 

For this work, magnifications were calculated from a ray-traced image library \citep{GravityCollaboration2018_orbital}.
The model is relativistic, albeit with a fast-light approximation \citep[for discusion see e.g.,][]{porth2019}. 
\autoref{fig:example_gr} illustrates the effects for two inclination angles. High inclination (i.e., edge-on view) leads to asymmetric peaks in the magnification and hence in the light curve, and the asymmetry shows up in a characteristic dependence of $\rm{SF_{\mu_3}(\tau)}$ on~$\tau$.  Low inclination leads to smaller, symmetric boosts in magnification, and $\rm{SF_{\mu_3}(\tau)}$ is near zero for all~$\tau$.

In order explore the relativistic effects, we constructed a  semi-physical model for Sgr~A*'s light curve:
\begin{equation}
    \begin{aligned}
    F_{\rm{model}}& (t, i, r_{\rm{f}}, \tau_{1,\rm{int}}, \tau_{2,\rm{int}}, \alpha, \beta, \kappa_{\rm{ou}}, \sigma_{\rm{ou}}, \theta_{\rm{ou}}) = \\ 
    &\sum_n a_n(\alpha, \beta)\cdot GR_n(t,i) \cdot C_{\rm{int}}(\tau_{1,\rm{int}}, \tau_{2,\rm{int}}) \\
    &+ AR_1(t, \kappa_{\rm{ou}}, \sigma_{\rm{ou}}, \theta_{\rm{ou}})  
    + \mathcal{N (\sigma_{\rm{obs}})}
    \end{aligned}
    \label{eq:gr_model}
\end{equation}
The first {part} in \autoref{eq:gr_model} represents a shot-noise model. This models the light curves as a sequence of flares, where the number of flares is drawn from a Poisson distribution with rate $r_{{f}}$. 
The flare amplitudes are drawn from an inverse gamma distribution\footnote{The choice of a gamma distribution was motivated by the observed flux-density distribution \cite{GravityCollaboration2020flux}.} with parameters $\alpha$, $\beta$. 
$GR_n(t,i)$ is the relativistic magnification kernel. To obtain the relativistic magnification for each flare, we assumed that the flares occur uniformly around the black hole with {distance from the black hole} drawn from a Gaussian distribution (mean $8R_g$, SD $1R_g$, bounded at [$6R_g$, $10R_g$]). The intrinsic emission of the flare was modeled with a double-exponential kernel function $C_{\rm{int}}(t, \tau_{1,\rm{int}},\tau_{2,\rm{int}})$ as in \autoref{eq: eq_kernel}, but the intrinsic rise and fall times in general will differ from the observed ones.
The second {part} in \autoref{eq:gr_model} is added to account for  low level variance in the data, typically referred to as the quiescence emission (e.g., \citealt{Dodds-Eden2009}). In order to model this quiescence emission, we used a generic time series model: the exponentiated Ohrnstein--Uhlenbeck ($\mathrm{ AR}_1$) process, which is identical to an first order autoregressive model ($\mathrm{AR}_1$). This process  models the overall data structure well and is by definition temporally symmetric \citep{Witzel2018}. 

Our model does not try to model the radiative emission mechanism. This is a limitation because the radiative process itself may well produce asymmetry in the light curve.  Examples include adiabatic expansion \citep[e.g.,][]{Yusef-Zadeh2009} and  radiative cooling \citep[e.g.,][]{aimar2023}. 
Nevertheless, the model constrains the asymmetry of the intrinsic emission (modeled by $C_{\rm{int}}[t,\tau_{\rm{1,int}},\tau_{\rm{2,int}}]$), though the intrinsic emission could be more asymmetric than the observed emission if the asymmetry induced by  relativistic effects happens to cancel the intrinsic asymmetry. 

The actual model fit, like the one in \autoref{sec:parameter_estimation}, was via the ABC algorithm. 

\subsection{Derived Sgr A* properties}
\DEL{
In this section, we detail the results of fitting the relativistic Sgr~A* model given in \autoref{eq:gr_model} to the observed light curves.
}

\begin{figure*}
    \centering
    \includegraphics[width=0.985\textwidth]{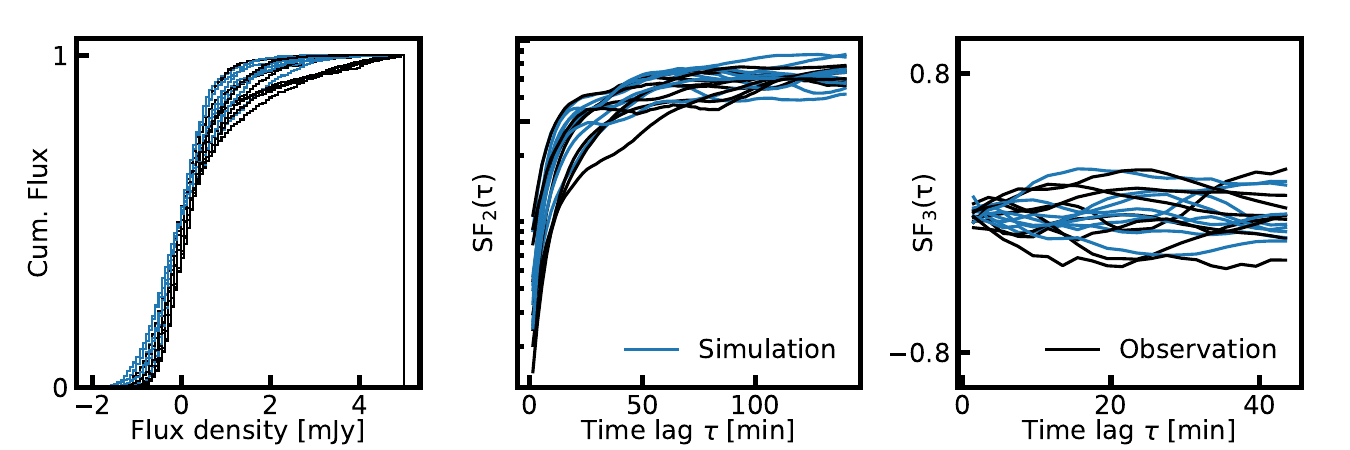}
    \caption{Summary statistics from posterior samples (blue) of the converged model fit compared to real observations (black). Statistics are shown separately for the eight observed light curves.  The left panel shows the cumulative flux-density distribution, and the other two panels show the two structure-function distributions as labeled.}
    \label{fig:results}
\end{figure*}

\begin{figure}
    \centering
    \includegraphics[width=0.485\textwidth]{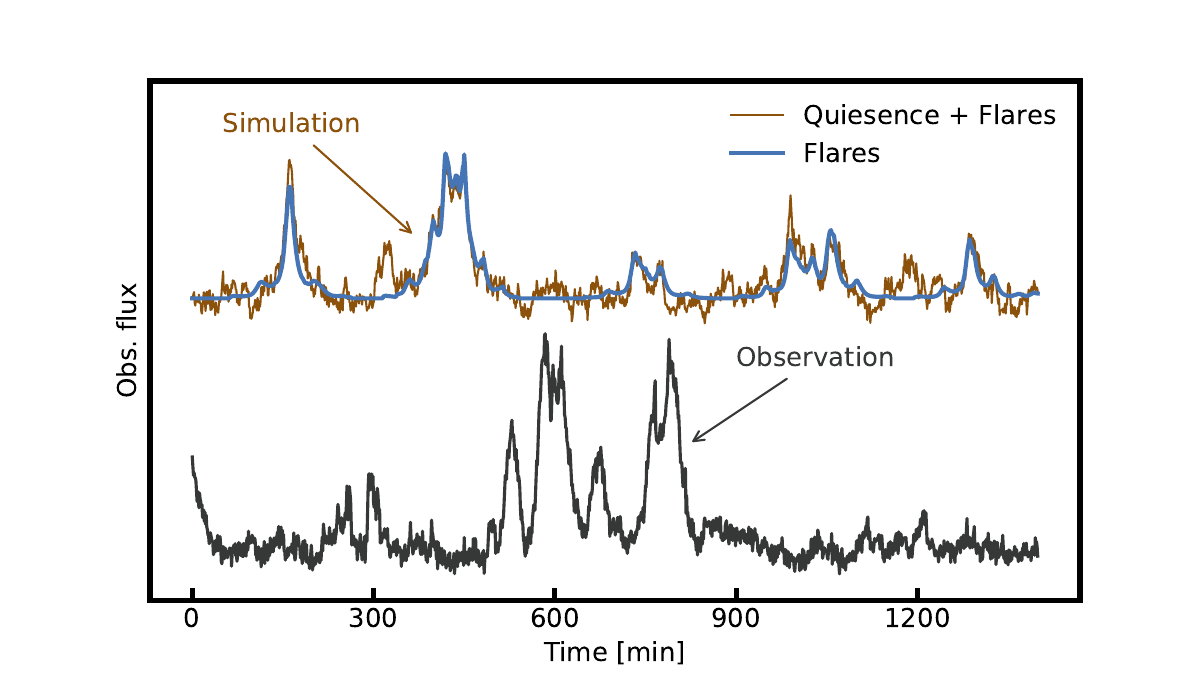}
    \caption{Simulated light curve draw from the model posterior compared to a {differential light curve measured by} Spitzer. The blue line shows simulated flares  (first {part} in \autoref{eq:gr_model} {)} drawn from the flare-model posterior. The brown line shows the full model light curve including the quiescence emission and observational noise, and the black line shows a Spitzer observation. The curves are vertically offset for clarity, but the  y-axis has the same linear scale for both curves.}
    \label{fig:post_example}
\end{figure}

The cumulative flux distribution, as well as the second- and third-moment structure function,  are well described by the model, as illustrated in \autoref{fig:results}. Light curves generated from the posterior distribution qualitatively match the observed light curves of Sgr A* (\autoref{fig:post_example}).

\begin{figure}
    \centering
    \includegraphics[width=0.485\textwidth]{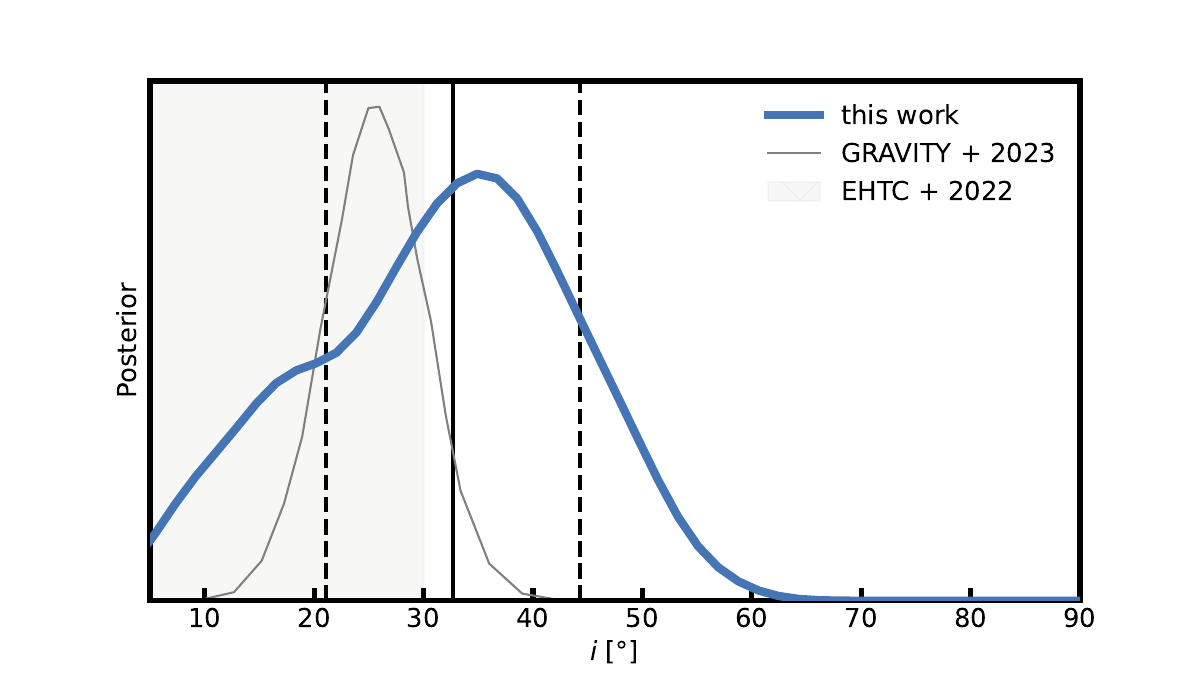}
    \caption{Inclination posterior of the relativistic model. Blue line shows the derived posterior, grey line shows the inclination posterior of \cite{GRAVITY_collab2023_polariflares}, and the gray shaded region shows the constraints derived from the 230~GHz image of the EHT Collaboration \cite{eht_sgra_I}.}
    \label{fig:inclination_dist}
\end{figure}

In the context of the model, high-inclination viewing angles are ruled out (\autoref{fig:inclination_dist}). This is because $\rm{SF_{\mu_3}}(\tau)$ shows no significant deviation from zero, inconsistent with the characteristic swing for high inclinations (\autoref{fig:example_gr}). 

Furthermore, the best fit shows no dip in $\rm{SF_{\mu_2}(\tau)}$, which would correspond to a peak in the power spectrum. This means the fit does not require significant periodicity, despite the flares being modeled as orbiting hot-spots. This is a consequence of the low flare rate and the different initial orbital positions. In addition, the preferred low inclinations cause only weak relativistic boosting and therefore little orbital modulation. Finding an orbiting-hot-spot model with no periodicity prove that the absence of significant periodicity is \textit{not} evidence against the orbiting-hot-spot model itself but is rather a consequence of the model parameters.

\begin{figure}
    \centering
    \includegraphics[width=0.485\textwidth]{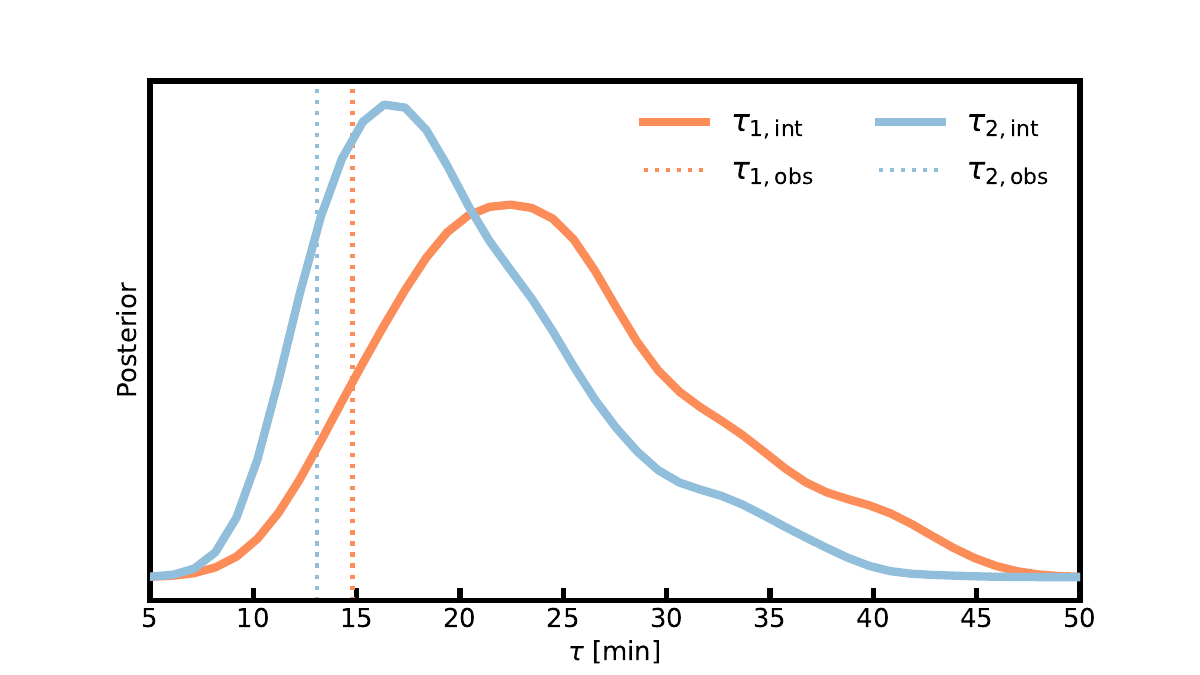}
    \caption{Posteriors of intrinsic time scales.  Colored curves show posteriors of $\tau_{1,\rm{int}}$  (orange) and $\tau_{2,\rm{int}}$ (blue) derived from fitting the relativistic model. Vertical dashed lines show the timescales ($\tau_{1,\rm{obs}}$ and $\tau_{2,\rm{obs}}$, colors as labeled) in the  observed light curves as derived from fitting the MA model.}
    \label{fig:timescales}
\end{figure}

The fit also {allows for} the intrinsic timescales to be longer than the observed time scales (\autoref{sec: ma_model}), as shown in \autoref{fig:timescales}. This  illustrates the ``speeding up'' of timescales by relativistic magnification. The speedup can also be quantified by the overall variance of the model light curves, which are altered by the magnification to show more and sharper peaks. This can be quantified by comparing posteriors with the same random seeds but relativistic effects set to either zero or one. The overall light curve variance is decreased by $\sim$20\% by the relativistic effects. 

\section{Discussion and Summary}
\label{sec: discussion}
This paper extends the results from Paper I, which showed that the average  shape of Sgr A*'s NIR and X-ray flares is symmetric. The present work confirms this symmetry in the entire light curve. The average observed rise and fall times are $\tau_{\rm{1,obs}} \approx \tau_{\rm{2,obs}} \approx 15~\mathrm{minutes}$. 

It is surprising that Sgr~A*'s overall light curve  shows no  significant asymmetry because the radiative processes thought  relevant for Sgr A* are typically asymmetric. For instance,  variability caused by electron-synchrotron cooling or adiabatic expansion after an initial acceleration event produces skewed light curves \citep[e.g.,][]{Dodds-Eden2009, aimar2023, Michail2024}. 

Paper I demonstrated that Doppler boosting of an orbiting hot spot can plausibly explain the symmetry in Sgr A*'s flares. While individual events may have strong asymmetry, depending on the starting position of the hot spot in the orbit, the boosting \textit{averaged} over multiple independent hot spots  will be symmetric. 
Paper~I also showed that hot spots in edge-on orbits produce asymmetric flares because of lensing. Such orbits are therefore inconsistent with the observed flare symmetry.
This paper corroborates these results by modeling the light curve with a semi-physical, relativistic model, composed of orbiting flares on top of a correlated red-noise-like quiescence emission.
The intrinsic process may be more asymmetric than the observed emission because of the averaging over orbits, 
but our model fits the observations only if  $i\la45\degr$ (\autoref{fig:inclination_dist}). This is consistent with other observations, which give stricter constraints.
 
Lastly, Sgr~A*'s light curves have a red-noise-like character with no significant periodicity. Nevertheless, our models show that orbiting hot spots in the accretion flow are consistent with these light curves. The absence of significant periodicity is  explained by a combination of  short intrinsic time scales, low inclinations, a low flare rate, and flares erupting at random orbit phases (i.e., at uniformly distributed orbit positions). 

\DEL{
\section{Summary}
\label{sec: conclusion}

This paper has introduced an easy but general definition of asymmetry in light curves.  The method itself is model-agnostic and insensitive to data selection. The idea is to generate mock  light curves via a random process acting on an underlying kernel having asymmetry as a parameter, then find the value of asymmetry for which the mock light curves best fit the observed data.
Using ABC to find the parameter values establishes the significance levels of the asymmetry. 

For Sgr~A*, light-curve analysis confirms the results from paper~I that, on average, X-ray and NIR flares are symmetric with  exponential rise and fall times $\tau\approx 15\mathrm{min}$. 
This is explained by the importance of Doppler boosting---but not lensing---in the light curve. The absence of strong lensing implies a viewing angle closer to face-on than edge-on.

To analyze the light curves, we introduced the third-moment structure function. When properly normalized, this function senses only the asymmetry of the light curve. Symmetric signals, such as white noise, average to zero.
}%
The third-moment structure function is a useful statistic for a wide variety of astronomical studies because many astronomical signals are temporally asymmetric.  Examples include astrometric signatures of orbiting black holes \citep{elbrady2023a, elbrady2023b}, light curves of eclipsing binaries \citep{ott1999, Gautam2024}, and even gravitational-wave signals \citep{ligoVirgoCatalog2019}. These examples illustrate the utility of the third-moment structure function well beyond the topic of this paper. 

\begin{acknowledgements}
{The authors would like to thank the referee, Dr.~Jeffrey~Scargle, for his very constructive and quick review, which significantly improved the manuscript.} H.-H. Chung is supported for this research by the International Max-Planck Research School (IMPRS) for Astronomy and Astrophysics at the University of Bonn and Cologne.
This publication is part of the M2FINDERS project which has received funding from the European Research Council (ERC) under the European Union’s Horizon 2020 Research and Innovation Programme (grant agreement No 101018682).

\end{acknowledgements}
%
\bibliographystyle{aa} 
\bibliography{bib2.bib} 
%

\appendix
\section{Properties of the third-moment structure function}\label{ap:SF3_properties}
\begin{figure*}
    \centering
    \includegraphics[width=0.99\textwidth]{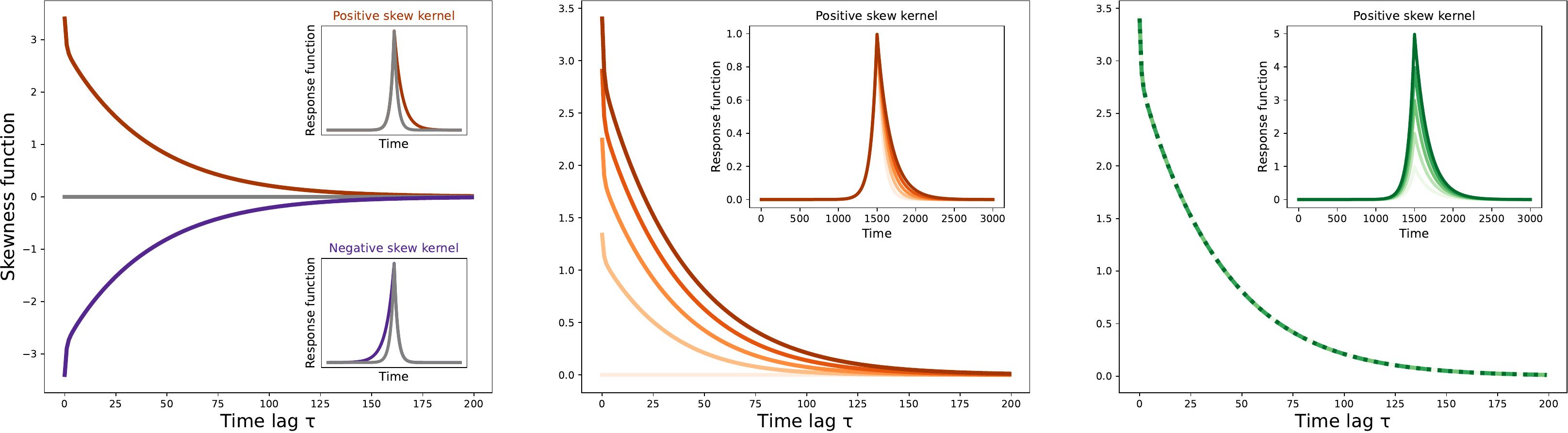}
    \caption{Third-moment structure functions $\mathrm{SF}_{\mu_3}$ for different kernels constructed from double-exponential profiles (\autoref{eq: eq_kernel}). Insets show example kernels, and the curves in the main panel show the resulting $\mathrm{SF}_{\mu_3}$.  \textit{Left:} Symmetric (grey) and asymmetric (positive in orange, negative in blue) kernels. \textit{Middle:} Five kernels with a fixed rise time  $\tau_1=15$ and fall times  $\tau_2=15$ (symmetric), 20, 25, 30, and 35. The larger $\mathrm{SF}_{\mu_3}$ values correspond to longer fall times as indicated by the curves' color saturation. \textit{Right:} Asymmetric kernels of different amplitudes. All have rise times  $\tau_1=15$ and fall times $\tau_2=30$, but amplitudes are multiplied by 1, 2, 3, 4, and 5. The structure function is the same for all amplitudes.}
    \label{fig: ma_kernel}
\end{figure*}

The third-moment structure function $\mathrm{SF}_{\mu_3}$ measures asymmetry.
\autoref{fig: ma_kernel} demonstrates that for an exponential impulse-response kernel, the structure function has a characteristic functional form. In particular, the larger the asymmetry (difference between rise and fall time), the stronger the signal in $\mathrm{SF}_{\mu_3}$. Because of the normalization, the structure function depends only on the time difference in the rise and decay times, not the amplitude of the kernel. 

\begin{figure*}[t!]
    \centering
    \includegraphics[width=1\textwidth]{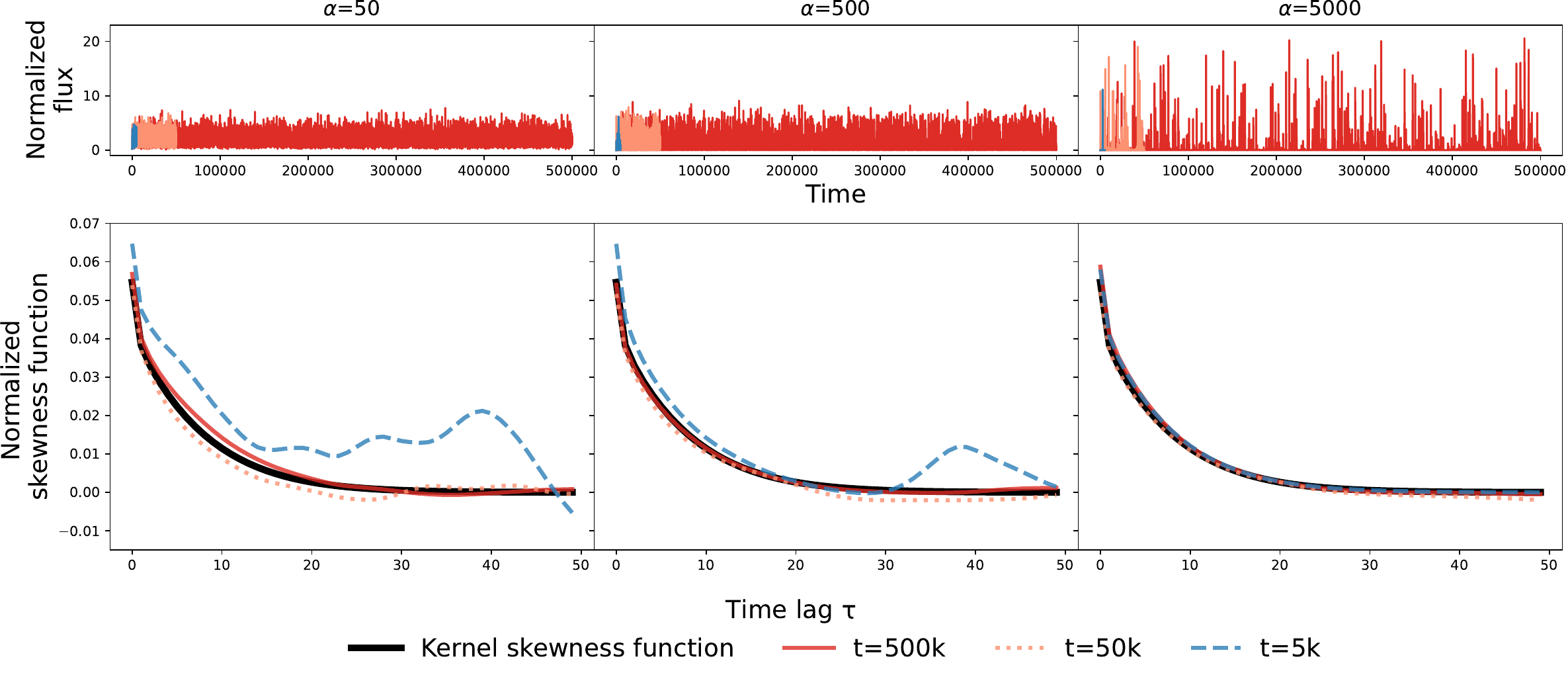}
    \caption{
    Simulated light curves and derived structure functions.
    \textit{Upper panels}:  Light curves for different values of $\alpha$ (\autoref{eq: innov}) as labeled.  
    The light curves were generated by the MA model (\autoref{eq: ma_process}) and include random and correlated noise with the random seed the same for all panels. The kernels are double-exponential profiles (\autoref{eq: eq_kernel}) having $\tau_1=?$ and $\tau_2=?$, and 
    the innovation $R=U^{\alpha}$ was drawn from uniformly distributed samples in [0, 1). Colors at the left mark sections of the light curve with 5\,000 and 50\,000 samples with the full curves containing 500\,000. 
    \textit{Lower panels}: Structure functions of the simulated light curve above each panel. 
    The black curve shows the true structure function without noise. The colored curves show the structure function derived from each light curve with a finite number of samples as labeled.
    }
    \label{fig: ma_process}
\end{figure*}

\autoref{fig: ma_process} demonstrates how the $\rm{SF}_{\mu_2}(\tau)$ function reacts to correlated noise. The three panels show light curve realizations generated with the same asymmetric impulse response and identical random seeds but different exponents $\alpha$ of the innovation (\autoref{eq: innov}). The longer the light curve, the better different $\rm{SF}_{\mu_3}(\tau)$ functions approximate the functional form of the skewness function of the impulse response. For shorter light curves, the errors are largest at large lag times, where there are fewer sample pairs, and fewer samples suffice when $\alpha$ is large because events are less blended.  Regardless of the correlation in the noise, if the sufficient data are available, the $\rm{SF}_{\mu_3}(\tau)$ function recovers the temporal asymmetry in the data. This demonstrates its utility: even correlated noise does not bias the results, especially at short lag times. 

\section{Models and fitting}
\begin{figure}
    \centering
    \includegraphics[width=0.45\textwidth]{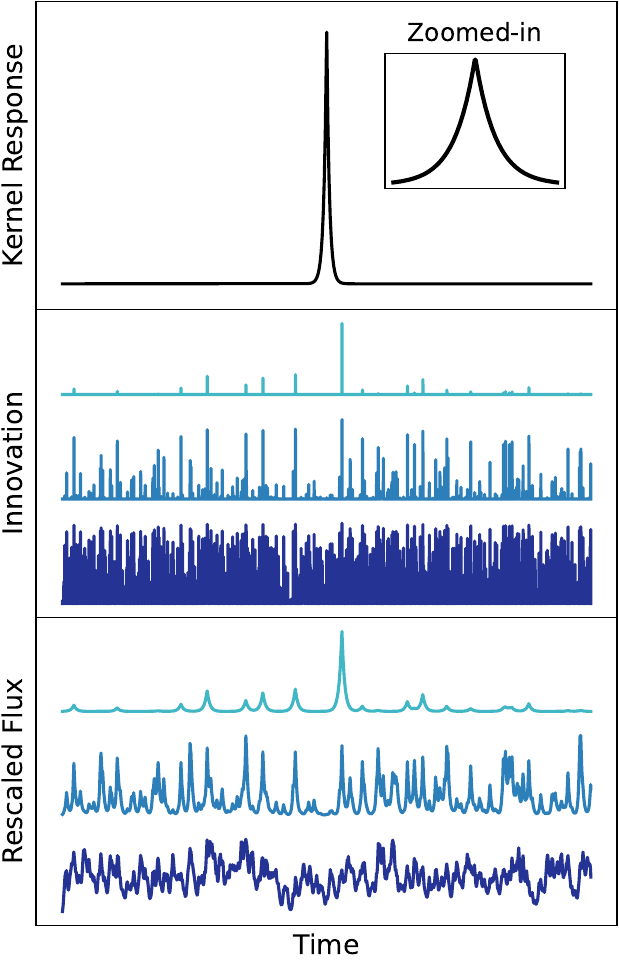}
    \caption{
    Simulated light curves generated based on the MA model. Top: temporally symmetric kernel response function.
    Middle: innovation $R(\alpha)$ for sparse, intermediate, and dense random impulses with $\alpha=1000$, 100, and 10, respectively.
    Bottom: the light-curve products convolved from the kernel and the corresponding innovation processes. Light curves are normalized by their maximal flux value for demonstration.
    }
    \label{fig: ma_lc_generation}
\end{figure}

\autoref{fig: ma_lc_generation} illustrates the properties of light curves generated from a moving average model. The figure shows the kernel function ($C(t,\tau_1, \tau_2)$ (impulse response) in the top panel. The middle panel shows three realizations of the innovation $\mathcal{U}^\alpha$ for $\alpha$=$1000$, $100$, and $10$ respectively. The higher the value of $\alpha$, the more the distribution is skewed. The bottom panel shows the light curves generated from such a process. The higher $\alpha$ the more ``isolated'' flares appear in the light curve, while for low $\alpha$ the concept of flare seems ill defined. Nevertheless, the underlying impulse response is the same.  

\subsection{Model parameters, priors, and posteriors}
We give model parameters, priors and posteriors in \autoref{tab: model_parameter}.
\begin{table*}
\renewcommand{\arraystretch}{1.5}
\centering
\caption{Model parameters, priors, and posteriors for the MA model and semi-physical Sgr A* model.}
\begin{tabular}{llcl} 
\hline\hline
parameters & prior & posterior & notes \\ 
\hline
MA model\\
$\alpha$  & uniform $\mathcal{U} \in [1,1000]$ & $218^{+14.3}_{-13.4}$ & exponent of innovation\\ 
$f_{\rm{scale}}$ & uniform  $\mathcal{U} \in [-100,100]$ &$22.4^{+1.3}_{-1.1}$ & flux density normalization (mJy)\\ 
$\tau_{1}$ & uniform $\mathcal{U} \in [1,100]$ & $14.8^{+0.4}_{-1.4}$ & rise time (minutes)\\ 
$\tau_{2}$ & uniform $\mathcal{U} \in [1,100]$ &  $13.8^{+1.3}_{-1.4}$ & fall time (minutes)\\ 
\hline
Sgr A*-model\\
$i$ & uniform $\mathcal{U} \in [5,90]$ & $32.7^{21.7}_{44.3}$ & inclination (degrees) \\
$\tau_{1, \rm{int}}$ & uniform $\mathcal{U} \in [10,35]$ & $23.6^{30.5}_{16.3}$ & intrinsic rise time (minutes)\\ 
$\tau_{2, \rm{int}}$ & uniform $\mathcal{U} \in [10,35]$ & $18.7^{24.8}_{12.6}$ & intrinsic fall time (minutes)\\ 
$r_{f}$ & uniform $\mathcal{U} \in [5, 300]$ & $10.1^{15.5}_{4.8}$ & flare rate parameter (Poisson)\\ 
$\mu_{\rm{amps}}$ & uniform $\mathcal{U} \in [0.1,1000]$ & $525^{738}_{313}$ & amplitude distribution parameter 1 (inv.\ $\gamma$)\\ 
$\sigma_{\rm{amps}}$ & uniform $\mathcal{U} \in [0.1,1000]]$ & $6259^{8549}_{3968}$ & amplitude distribution parameter 2 (inv.\ $\gamma$)\\ 
$\sigma_{\rm{measurement}}$  & uniform $\mathcal{U} \in [0.05,0.6]]$ & $0.16^{0.20}_{0.12}$ & measurement noise (mJy)\\ 
$\kappa_{\rm{O.U.}}$ & uniform $\mathcal{U} \in [0.01,0.1]]$ & $0.06^{0.08}_{0.05}$ & exp.\ Ornstein Uhlenbeck parameter 1\\ 
$\theta_{\rm{O.U.}}$ & uniform $\mathcal{U} \in [0.1,0.4]]$ & $0.6^{0.9}_{0.35}$ & exp.\ Ornstein Uhlenbeck parameter 2\\ 
$\sigma_{\rm{O.U.}}$ & uniform $\mathcal{U} \in [0.001,0.1]]$ & $0.08^{0.10}_{0.07}$ & exp.\ Ornstein Uhlenbeck parameter 3\\ 
\hline
\end{tabular}
\label{tab: model_parameter}
\end{table*}

\subsection{Distance functions and model convergences}
We used three distance function for the likelihood interference scheme. The second moment structure function of the simulated data was compared to the observed data using a distance function akin to the $\log \chi^2$ function:
\begin{equation}
    \log \chi^2 = 0.5 \left( \log \rm{SF}_{\rm{obs_i,}~\mu_2} / \log\rm{SF}_{\rm{sim_i,}~ \mu_{2}} \right)^2.
\end{equation}
The third moment structure function of the simulated data was compared to observed data using a distance function akin to the $\chi^2$ function:
\begin{equation}
    \chi^2 = 0.5 \left( \rm{SF}_{\rm{obs_i,}~\mu_2}  - \rm{SF}_{\rm{sim_i,}~ \mu_{2}} \right)^2,
\end{equation}
which allows to account for the factor that the third moment structure function can have positive as well as negative values. 
{The time lags were binned using $3~\mathrm{min}$ linear time bins, which was the empirically determined optimum between time sampling and noise in the structure functions.}

For the MA model we used to the classical KS-test to compare observed and simulated flux distribution, which turned out to be too insensitive to tail outliers for the semi-physical Sgr~A*. For the latter, we used cumulative (flux distribution) histogram fitting as a distance function, where we calculated the $\chi^2$ distance between the histogram's bins. 

We employ the Population Monte Carlo (PMC) algorithm \citep{Beaumont_2010} implemented in abcpy, which converges using `populations' of random samples, which are iteratively generated, with each generation posing tight constraints on the parameters. One practical convergence critirum is that more generations do not lead to substainly different constraints on the derived parameters. We illustrate this is the case for both the generic MA model, as well as for the semi-physical relativistic Sgr A* in the following.
\subsubsection{MA model}
\begin{figure}
    \centering
    \includegraphics[width=0.485\textwidth]{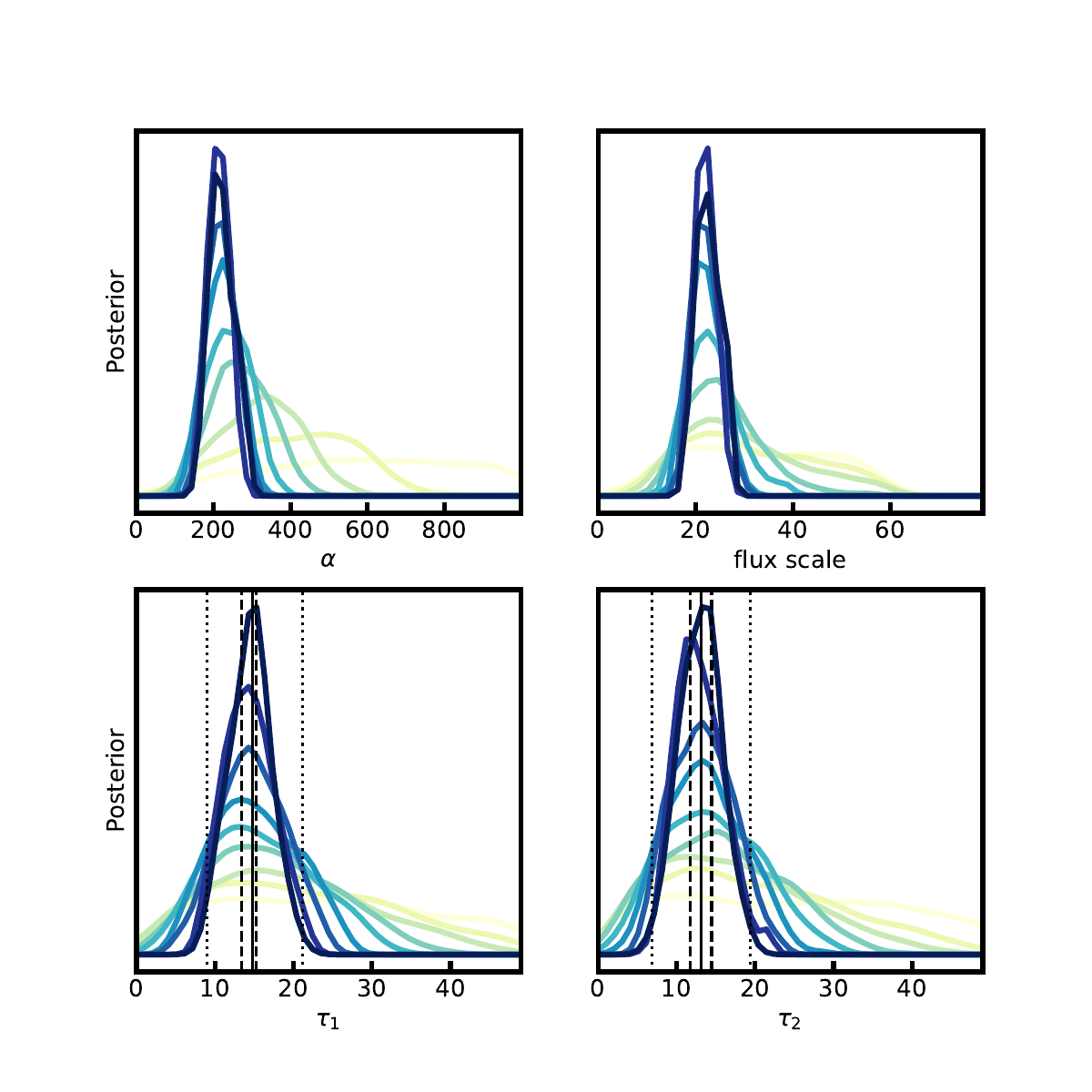}
    \caption{Posterior for the model (\autoref{eq:model}) generated using the Approximate Bayesian Computation (ABC) method. The different colored lines indicate different ``sample generations'', with darker colors representing more converged chains. }
    \label{fig:convergences_mamodel}
\end{figure}

\autoref{fig:convergences_mamodel} illustrates that we have run the abcpy's PMC suffiecently long to derive meaning full parameter constraints. The quantilies of the distribution do not change substainly between the last generation and the second last.
\subsubsection{Relativistic Sgr A* model}
\begin{figure*}
    \centering
    \includegraphics[width=0.985\textwidth]{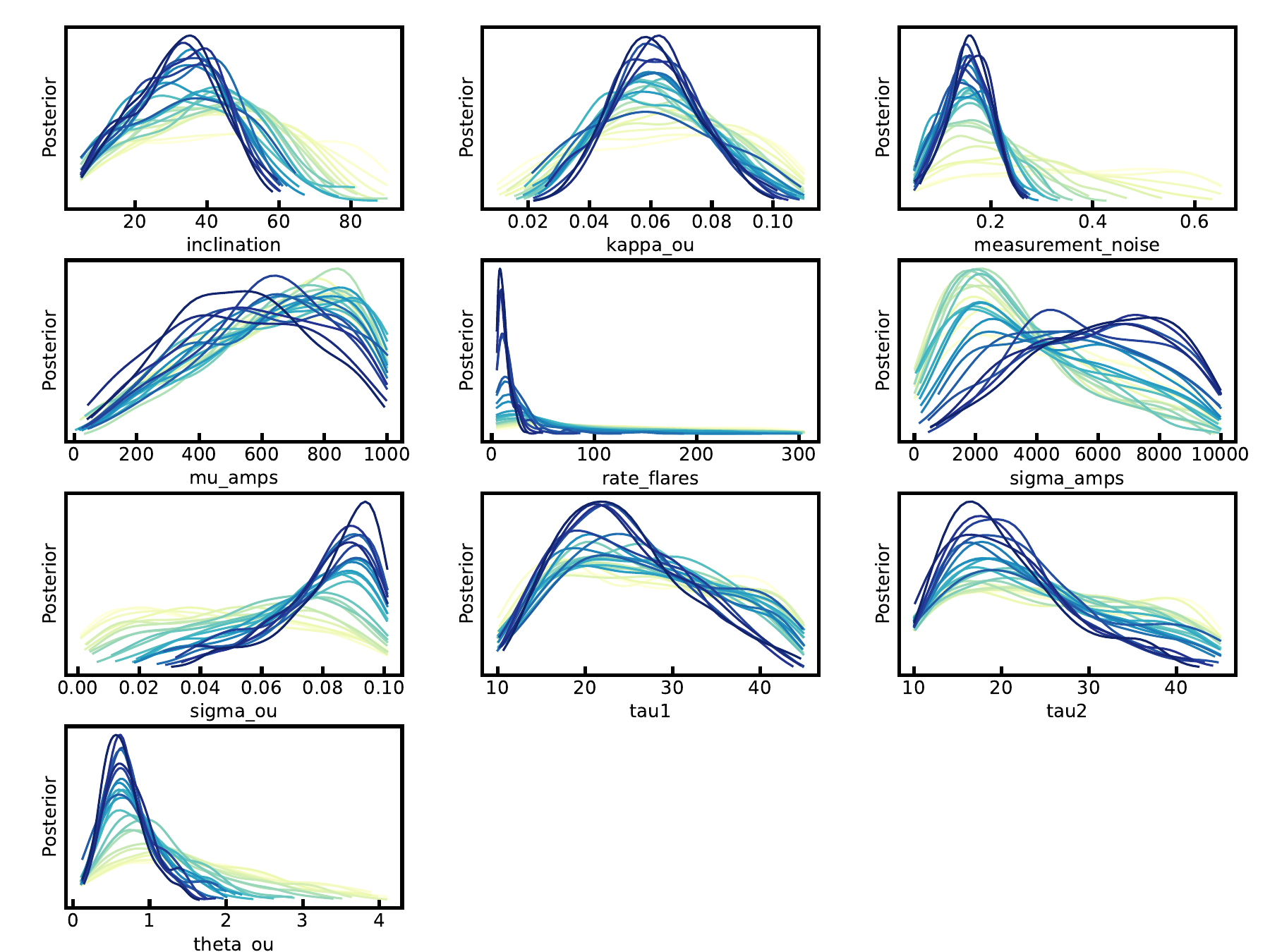}
    \caption{Similar as \autoref{fig:convergences_mamodel}, but for the convergences behavior of the of the relativistic Sgr A* model. Dark colors indicate later populations.}
    \label{fig:convergences_sgramodel}
\end{figure*}

As for the MA model, \autoref{fig:convergences_sgramodel} illustrates that the alogrithm has run for sufficient  time to be considered converged.

\section{Proof of $\rm{SF}_{\mu_3}$ asymmetry properties} \label{ap:proof}

We would like to show that for a temporally symmetric light curve $\rm{SF_{\mu_3}}$ is zero.

Let $\{F_1(t)\}$ be a stationary random process with $t=0, \pm 1, \pm 2, ...$ and $\{F_2(t)\}$ as second stationary random process with $F_2(-t) = F_1(t)$ for all $t$. Let us define $\rm{SF}_{\mu_3}(\tau)$ and $\widehat{\rm{SF}}_{\mu_3}(\tau)$  to be the respective structure functions (see definition \autoref{eq: SF}).

Then, 
    \begin{align*}
    &-\widehat{\rm{SF}}_{\mu_3} (\tau)= - 1/N_i \sum_{t_j,t_i} (F_2(t_{i}) - F_2(t_{j}))^3 \\ 
                        & = 1/N_i \sum_{t_j,t_i} (F_2(t_{j}) - F_2(t_{i}))^3 \\
                        &= 1/N_i \sum_{t_j,t_i} (F_1(t_{i}) - F_1(t_{j}))^3  = \rm{SF}_{\mu_3}(\tau), \\         
    \end{align*}

i.e., time reversal leads to a change of sign of the third moment structure function.

In this paper, we define temporal symmetry not in the sense of a deterministic function (i.e., $F(t) = F(-t)$), but in a statistical sense. This can be achieved by adopting a definition similar to the definition of stationarity, as given in \cite{Priestley1988}:

\textbf{Definition.} The random process $\{F_1(t)\}$ is said to be statistically time-symmetric up to order m if, for any admissible $t_1, t_2, ..., t_n$ all the joint moments up to order $m$ of $\{F_1(t_1), F_1(t_2), ..., F_1(t_n)\}$ exist and equal the corresponding joint moments up to order m of $\{F_2(t_1), F_2(t_2), ..., F_2(t_n)\}$, with $F_2(t) = F_1(-t)$ for all $t$.

Then also all the corresponding moments on time lag selected differences are equal, e.g., $\rm{SF}_{\mu_3}$ = $\widehat{\rm{SF}}_{\mu_3}$.

If the third order structural functions are equal and negative to one another, their values have to be zero.

\end{document}